\documentclass{ecmtb2002}        %! needed for authors LM/LSG

\usepackage[dvips]{epsfig}       %! needed for authors LM/LSG
\usepackage{exscale}
\usepackage[dvips]{color}
\usepackage{latexsym}
\usepackage{amssymb}             %! needed for authors LM/LSG
\usepackage{amsmath}             %! needed for authors LM/LSG
\usepackage{makeidx}
\makeindex
\begin{document}

\title*{A computational approach to regulatory element discovery in eukaryotes}
\toctitle{A computational approach to regulatory element discovery
in eukaryotes}

\titlerunning{Regulatory element discovery
in Eukaryotes}

\author{Michele Caselle \inst{1}
\and Ferdinando Di Cunto\inst{2}
\and Paolo Provero\inst{3}}
\authorrunning{M. Caselle, F. Di Cunto, P. Provero}
\institute
{Dip. di Fisica Teorica, Universit\`a di Torino and INFN, sez. di
Torino, Italy
\and
Dip. di Genetica, Biologia e Biochimica, Universit\`a di Torino, Italy
\and
Dip. di Scienze e Tecnologie Avanzate, Universit\`a del Piemonte
Orientale and INFN, gruppo collegato di Alessandria, Italy}

\maketitle              % typesets the title of the contribution

%\begin{document}

\index{Caselle, M.}

\begin{abstract}
Gene regulation in Eukaryotes is mainly effected through
transcription factors binding to rather short recognition motifs
generally located upstream of the coding region. We present a
novel computational method to identify regulatory elements in the
upstream region of Eukaryotic genes. The genes are grouped in sets
sharing an overrepresented short motif in their upstream sequence.
For each set, the average expression level from a microarray
experiment is determined: if this level is significantly higher or
lower than the average taken over the whole genome, then the
overrepresented motif shared by the genes in the set is likely to
play a role in their regulation. We illustrate the method by
applying it to the genome of {\it S. cerevisiae}, for which many
datasets of microarray experiments are publicly available. Several
known binding motifs are correctly recognized by our algorithm,
and a new candidate is suggested for experimental verification.
\end{abstract}
\section{Introduction}
\label{sec:intro} One of the most interesting subjects of
computational genomics is the characterization of the network of
regulatory interactions between genes. A dominant role in this web
of interactions is played by the mechanism of transcriptional
regulation, in which one or more gene products regulate the
transcription of other genes into mRNA, by binding to recognition
motifs (short DNA sequences) generally located upstream of the
coding region of the regulated gene. A key step in the study of
transcriptional regulation is the identification of such
recognition motifs.
\par
In recent years, it has become possible to attack this problem on
a genomic scale with computational methods, thanks to the huge
amount of data generated by modern experimental techniques. In
particular, two sets of experimental data are of utmost importance
to the study of transcriptional regulation: first, the
availability of the complete, annotated DNA sequence for many
model organisms allows the systematic exploration of the upstream
regions where the motifs are to be found; second, microrarray
techniques allow the identification of groups of coregulated
genes.
\par
When trying to identify regulatory motifs within a computational
approach, one is interested in finding instances of statistical
correlation between the presence of certain sequences in the
upstream region and common patterns of expression. This search is
facilitated by the fact that, in many cases, the regulatory motif
must be repeated several times in the upstream region to be
effective, and therefore the number of its appearances in the
upstream region is above what expected from chance alone. The most
common approach to the problem consists in two steps: first, a set
of co-expressed genes is identified, either using direct
experimental knowledge or by applying clustering techniques to
microarray data; then the upstream region of such genes are
systematically analyzed, looking for motifs that are
overrepresented in some statistically precise sense. Examples of
this procedure as applied to yeast ({\it Saccharomyces
cerevisiae}) are Refs.
\cite{DeRisi:1997,vanHelden:1998,Tavazoie:1999}.
\par
In this work we propose an alternative procedure, which in some
sense reverses the procedure: the genes are grouped based on the
short sequences that are overrepresented in their upstream region,
and such groups are then tested for evidence of coregulation,
using microarray data. Whenever such evidence of coregulation is
found, the sequence characterizing the upstream region of the
genes is a candidate regulatory site. 
The most attractive feature of this reversed procedure is that the
grouping of the genes is based on the sequence only:  
the sets of genes thus obtained can be analysed for evidence of
coregulation not only by using microarray data, as 
described in this contribution, but also by comparing them to other
biological data, such as {\it e.g.} functional annotation and
the composition of protein complexes.
\par
A somewhat related method,
that does not rely on previous clustering of the genes, is
described in Ref. \cite{Bussemaker:2001}. In this contribution we
focus on describing the method: for a detailed discussion of the
results and comparison with other methods, we refer the interested
reader to the original paper, Ref.\cite{Caselle:2002}.
\section{Grouping the genes based on overrepresented upstream motifs}
\label{section:group} The first step in our procedure consists in
analyzing the upstream region of all the $\sim$6000 yeast genes,
and identify for each gene the short DNA sequences ({\it words})
which are overrepresented in their upstream region.
Overrepresentation of a word in the upstream sequence of a gene is
defined by comparison to the prevalence of the same word in all
the upstream regions taken as a single sample.
\par
Specifically, for each possible word $w$ (of length 6-8 in the
present study) we construct the set $S(w)$ of the genes in whose
upstream region $w$ is overrepresented by the following procedure:
\begin{itemize}
\item
For each yeast gene $g$ we compute the number $m_g(w)$ of
occurrences of $w$ in the upstream region of $g$. The length of
the upstream region we consider is 500 base pairs, but is
shortened whenever necessary to avoid overlapping with the coding
region of the preceding open reading frame (ORF). 
Therefore the length of the coding
region depends on the gene considered, and will be denoted by
$K_g$. Non palindromic words are counted on both strands:
therefore we define the effective number of occurrences  $n_g(w)$
as
\begin{eqnarray}
n_g(w)&=&m_g(w)+m_g(\tilde{w}) \qquad {\rm if}\  w\ne\tilde{w}\\
n_g(w)&=&m_g(w)\qquad {\rm if}\  w=\tilde{w}
\end{eqnarray}
where $\tilde{w}$ is the reverse complement of $w$.
\item
 We define the global frequency $p(w)$ of each word $w$ as
 \begin{equation}
 p(w)=\frac{\sum_g n_g(w)}{\sum_g L_g(w)}
 \end{equation}
where, in order to count correctly the available space for
palindromic and non palindromic words,
 \begin{eqnarray}
L_g(w)&=&{2(K_g-l+1)}\qquad {\rm if}\  w\ne\tilde{w}\\
L_g(w)&=&{(K_g-l+1)}\qquad {\rm if}\  w=\tilde{w}
\end{eqnarray}
where $l$ is the length of $w$. $p(w)$ is therefore the frequency
with which the word $w$ appears in the upstream regions of the
whole genome: it is the ``background frequency'' against which
occurrences in the upstream regions of the individual genes are
compared to determine which words are overrepresented.
\item
For each ORF $g$ and each word $w$ we compute the probability
$b_g(w)$ of finding $n_g(w)$ or more occurrences of $w$ based on
the global frequency $p(w)$:
\begin{equation}
b_g(w)=\sum_{n=n_g(w)}^{L_g(w)} \left( {L_g(w)}\atop {n}\right)
p(w)^n \left[ 1-p(w)\right]^{L_g(w)-n}
\end{equation}
\item
We define a maximum probability $P$ and consider, for each $w$,
the set
\begin{equation}
S(w)=\{ g:\ b_g(w)<P\}
\end{equation}
of the ORFs in which the word $w$ is overrepresented compared to
the frequency of $w$ in the upstream regions of the whole genome.
That is, $w$ is considered overrepresented in the upstream region
of $g$ if the probability of finding  $n_g(w)$ or more instances
of $w$ based on the global frequency is less than $P$. In this
study we chose $P=0.02$. Note that this is a rather lenient
cutoff: however no biological significance is attributed to the
set $S(w)$ in itself: only those sets which will pass the test for
coregulation, described in the following section, will be
considered of biological relevance, and hence it is on this second
test that a stringent cutoff on P-value must be imposed.
\end{itemize}
\section{Looking for coregulation within the sets $S(w)$}
\label{section:coreg} The second step of the procedure consists in
verifying whether the genes contained in each set $S(w)$ show
evidence of coregulation in their expression levels as measured in
microarray experiments. This is done by computing the average
expression  level of the genes in the set and comparing it to the
average expression level for the same experiment computed over the
whole genome. We used the data from the diauxic shift experiment
of Ref.\cite{DeRisi:1997} which provides us with expression
measurements for virtually all the yeast genes at 7 timepoints
corresponding to progressive depletion of the glucose in the
medium, with the consequent metabolic shift form fermentation to
respiration. We proceed as follows:
\begin{itemize}
\item
For each time-point $i$ we computed the genome-wide average
expression $R(i)$ and its standard deviation $\sigma(i)$.
\item
Then for each word $w$ we compute the average expression in the
subset of $S_w$ given by the genes for which an experimental
result is available at timepoint $i$ (in most cases this coincides
with $S_w$):
\begin{equation}
R_w(i)=\frac{1}{N(i,w)} \sum_{g\in S_w} r_g(i)
\end{equation}
where $N(i,w)$ is the number of ORFs in $S_w$ for which an
experimental result at timepoint $i$ is available, and $r_g(i)$ is
the corresponding expression level (as customary, this is defined
as the $\log_2$ of the ratio between the mRNA level at timepoint
$i$ and the initial mRNA level).
\item
The difference
\begin{equation}
\Delta R_w(i)=R_w(i)-R(i)
\end{equation}
represents the discrepancy between the genome-wide average
expression at time-point $i$ and the average expression at the
same time-point of the ORFs that share an abundance of the word
$w$ in their upstream region.
\item
 A significance index ${\rm sig}(i,w)$ is
defined as
\begin{equation}
{\rm sig}(i,w)=\frac{\Delta R_w(i)}{\sigma(i)}\sqrt{N(i,w)}
\label{sig}
\end{equation}
and the word $w$ is considered significantly correlated with
expression at time point $i$ if
\begin{equation}
|{\rm sig}(i,w)|>\Lambda
\end{equation}
In this work we chose $\Lambda=6$: this means that we consider
meaningful a deviation of $R_w(i)$ by six s.d.'s from its expected
value. The sign of ${\rm sig}(i,w)$ indicates whether $w$ acts as
an enhancer or an inhibitor of gene expression.
\end{itemize}
\section{Results}
\label{section:res} A total of 29 words of length between 6 and 8
base pairs pass our significativity test for at least one
timepoint in the diauxic shift experiment.  Two of them must be
eliminated as their statistical significance can be ascribed to
the existence of a family of several nearly identical ORFs with
nearly identical upstream regions. Out of the remaining 27 words,
26 can be confidently identified with a known regulatory motif,
and one word is a new candidate. Examples of significant words,
including the new candidate ATAAGGG, are reported in Tab.
\ref{tab:results}.
\begin{table}
\centering \caption{Example of regulatory words found by our
method, together with the known regulatory motif they belong to.
In the third column we report the timepoint(s) of the diauxic
shift experiment where the word passes the significativity test.
The sign in the fourth column indicates whether the genes in the
set $S(w)$ are induced or repressed at these timepoints.}
\label{tab:results}       % Give a unique label
\begin{tabular}{llll}
\hline\noalign{\smallskip}
word& motif &timepoints& +/-   \\
\noalign{\smallskip}\hline\noalign{\smallskip}
GATGAG & PAC & 4 & - \\
GATGAGAT & PAC & 4,5,6,7& -\\
GATGAGA & PAC & 4,7& -\\
AAAATTT & RRPE & 6,7&-\\
AAAATTTT & RRPE & 4,6,7 &-\\
CCACCCCC & STRE & 6 &+\\
CCCCCCCT & STRE & 6 & +\\
TACCCC & MIG1 & 6 & +\\
GCCGCC & UME6 &7 &+\\
ATAAGGG &{\bf new} &6,7&+\\
\noalign{\smallskip}\hline
\end{tabular}
\end{table}

\begin{figure}
\centering
\includegraphics[height=4cm]{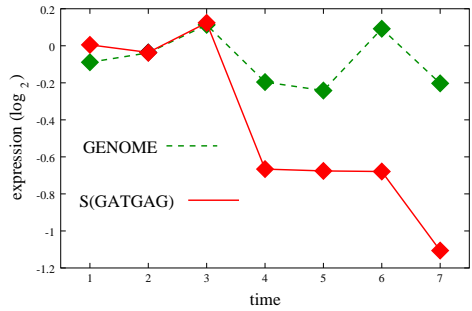}
\caption{Expression of the genes in the set S(GATGAG): The average
expression of the genes in the set (solid line) is compared to the
genome-wide average expression (dashed line) at the seven time
points of the diauxic shift experiment.}
\label{figure:gatgag1}       % Give a unique label
\end{figure}

\begin{figure}
\centering
\includegraphics[height=4cm]{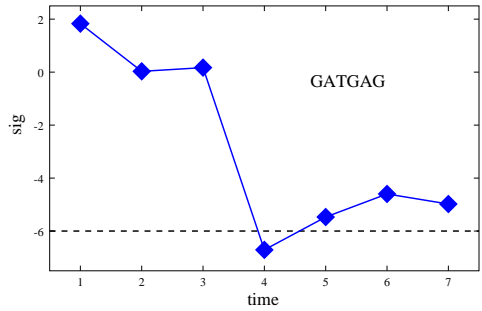}
\caption{Statistical significance ${\rm sig}(i,w)$ as defined in
Eq.(\ref{sig}) for the word $w={\rm GATGAG}$ and timepoints
$i=1,\dots,7$ in the diauxic shift experiment. The dashed line is
the significance threshold $|{\rm sig}|=6$.}
\label{figure:gatgag2}       % Give a unique label
\end{figure}

In Figs. \ref{figure:gatgag1} and \ref{figure:gatgag2} , we give
an example of coregulation in the set $S(w)$ corresponding to the
significant word GATGAG

\section{Conclusions}
\label{section:concl} Our method appears to be very efficient in
identifying regulatory motifs whose effectiveness depend on many
repetitions of a short sequence in the upstream region of the
regulated genes.  The main feature that differentiates our method
from existing algorithms for motif  discovery is the fact that
genes are grouped {\it a priori} based on similarities in their
upstream sequences. 
The fact that most of the motif revealed by
the method turn out to be known regulatory motifs suggests a very
low rate of false positives.
Also in Ref.\cite{Bussemaker:2001} a computational method was proposed
in which candidate binding sites are identified using statistical
correlations between upstream sequence and expression data. Two important
differences can be identified between our method and the one proposed
in Ref.\cite{Bussemaker:2001} (see Ref.\cite{Caselle:2002} for a
detailed comparison of the results): first, our
algorithm does not assume 
a linear dependence of the expression level from the number of
repetitions of the motif. Second, our sets of genes, being based on
the upstream sequence only, can be screened for evidence of coregulation not
only by comparison with expression data, as described above, but also
with other types of biological datasets, such as functional annotation
or experimentally determined protein complexes.
% For tables use

% For LaTeX tables use
%
%

% For figures use
%

% Non-BibTeX users please use

\end{document}